\documentclass[prd,aps,nofootinbib,preprint,eqsecnum]{revtex4}

\usepackage{hhline}
\usepackage{latexsym,graphicx}
\usepackage{subfigure}
\usepackage{tikz}
\usepackage[percent]{overpic}
\usepackage{pgfplots}
\pgfplotsset{compat=1.18}
\usepackage{amssymb}
\usepackage{amsmath}
\usepackage{amscd}
\usepackage{amsthm}
\usepackage{float}
\usepackage{xcolor}
\usepackage{comment}

\begin{document}
\title{Non-equilibrium thermodynamics of an expanding Swiss-Cheese braneworld cosmology with a matter bounce}

\author{Nasr Ahmed$^1$}
\email[]{nasr.ahmed@nriag.sci.eg}
\affiliation{$^1$ Astronomy Department, National Research Institute of Astronomy and Geophysics, Egypt.}

\begin{abstract}

We investigate the non-equilibrium thermodynamics of a Swiss-cheese (SC) braneworld universe. By employing Hayward's unified first law and the Clausius relation at the apparent horizon, we first show that the standard equilibrium treatment of the effective SC braneworld fluid, together with the usual Hawking temperature, reproduces the Bekenstein-Hawking area law without any braneworld correction to the entropy. This demonstrates that the brane quadratic energy density corrections cannot, by themselves, generate a modified entropy area relation within the equilibrium framework. We therefore formulate a non-equilibrium thermodynamic description in which deviations from the Bekenstein-Hawking entropy are accompanied by an internal entropy production term. For a general horizon entropy functional, we derive the corresponding entropy production rate and then specialize to a logarithmic and inverse area quantum-corrected entropy. The resulting irreversible contribution is expressed in terms of the Hubble parameter and its derivative. We apply the formalism to a nonsingular matter-bounce solution in the SC braneworld. During the expanding phase, the total entropy production remains positive satisfying the GSLT. Moreover, the late-time behavior of the second derivative of the total entropy becomes negative, indicating a decelerating entropy growth and an asymptotic approach toward thermodynamic equilibrium. Far from the bounce, the equation of state parameter $w \approx -1$, indicating a dark energy-dominated regime. Significant departures from $w = -1$ are confined to the bounce vicinity, where high energy braneworld corrections drive the nonsingular transition. Thus, the model naturally transitions from a high energy bounce into an asymptotic dark energy phase during late-time expansion.

\end{abstract}
\maketitle

\section{Introduction}
 The discovery of Hawking radiation \cite{hawkk} established a profound connection between gravity and thermodynamics by demonstrating that black hole entropy $S$ is directly proportional to its horizon area $A$ \cite{therm}. Within black hole thermodynamics, the first law $T dS = dE$ links the horizon entropy $S = \frac{A}{4G}$ with the Hawking temperature $T = \frac{\vert{}\kappa_{sg}\vert{}}{2\pi}$, where $\kappa_{sg}$ is the surface gravity and $dE$ denotes the energy change \cite{61,62,63}. More generally, the energy variation takes the form $dE = T dS + \text{work terms}$, where the work terms depend on the geometry of the underlying black hole. This framework provides a thermodynamic basis for the Einstein field equations near the horizon, mapping geometric attributes of black hole solutions directly to thermodynamic parameters \cite{80,81}.Jacobson significantly advanced this paradigm \cite{64} by showing that the Einstein equations emerge naturally from the Clausius relation ($T dS = \delta Q$) alongside the horizon area law, where $\delta Q$ represents the energy flux across the horizon and $T$ is the Unruh temperature. Hayward later extended this formulation to spherically symmetric spacetimes \cite{82} by deriving the unified first law of black hole dynamics and relativistic thermodynamics, $dE = T dS + W dV$, with the work density defined as $W = -\frac{1}{2} T^{ab} h_{ab} = \frac{1}{2}(\rho - p)$.

\quad  Applying the Clausius relation at the apparent horizon of an FLRW universe enables a thermodynamic derivation of the standard cosmological Friedmann equations \cite{65, bo, boo}. As demonstrated in \cite{60}, these equations take the form $dE = T dS + W dV$ at the horizon boundary, where $E = \rho V$ is the total energy and $W = \frac{1}{2}(\rho - p)$ is the work density \cite{82}. Here, $T$ and $S$ denote the temperature and entropy associated with the apparent horizon. The standard area law $S \propto A$ holds strictly within General Relativity Consequently, it must be modified when considering theories with higher-order curvature terms \cite{basic1}. This motivates the question of whether the modified cosmological equations can still be recovered from generalized entropy-area relations. A possible answer has been suggested in Ref. \cite{basic1} where the authors successfully derived the modified equations by applying the quantum-corrected entropy 
\begin{equation} \label{ent}
S=\frac{A}{4G}+\alpha \ln \frac{A}{4G}+\beta \frac{4G}{A}.
\end{equation}
where $\alpha$ and $\beta$ are dimensionless constants. The accurate values of these parameters still controversial and are not fully fixed even within loop quantum gravity \cite{salehi}. In general, the correction terms in Eq. (\ref{ent}) originate from quantum effects, thermal equilibrium, or mass fluctuations \cite{echde}. Although existing literature presents arguments for both positive and negative choices for $\alpha$ and $\beta$ \cite{zer1}-\cite{zer5}, it has also been suggested in \cite{zer} that the logarithmic prefactor may most plausibly vanish. 

\quad According to the maximum entropy principle, a universe reaching thermodynamic equilibrium at late times must satisfy $\dot{S} \geq 0$ and $\ddot{S} \leq 0$, ensuring that its total entropy grows monotonically toward a finite upper bound \cite{entropy1, entropy2}. The first inequality guarantees that the apparent horizon entropy is non-decreasing, while the second enforces the approach toward equilibrium, where entropy reaches its peak with no more increase. Furthermore, the GSLT dictates that the sum of the matter entropy within the apparent horizon and the horizon's own entropy must satisfy $\dot{S}_m + \tilde{\dot{S}}_h \geq 0$.

\section{Swiss-Cheese Brane-worlds}

\quad In the brane-world scenario—first introduced via the domain wall concept \cite{domain}—our $(3+1)$-dimensional universe acts as a hypersurface embedded in a $(3+1+N)$-dimensional bulk. Various formulations exist, such as universal extra dimension models \cite{yours}, the Dvali–Gabadadze–Porrati (DGP) \cite{brane6ii}, Gregory-Rubakov-Sibiryakov (GRS) \cite{brane6i}, and thick brane scenarios \cite{brane6iii}. Brane-world cosmology introduces a quadratic energy density term ($\rho^2$) that alters inflationary dynamics and reshapes the initial Big Bang singularity in anisotropic spacetimes \cite{ukra}. The Einstein equations on the brane are given by \cite{eins} 
\begin{equation}
G_{ab}=-\Lambda g_{ab}+\kappa^2T_{ab}+\tilde{\kappa}^4S_{ab}-\varepsilon_{ab},
\end{equation}

Here, $G_{ab}$ and $g_{ab}$ represent the Einstein and metric tensors respectively, while $\varepsilon_{ab}$ denotes the electric component of the bulk Weyl tensor. The tensor $S_{ab}$ captures the quadratic energy-momentum contributions and is defined as:
\begin{equation}
S_{ab}=\frac{1}{12}TT_{ab}-\frac{1}{4}T_{ac}T^c_{\ b}+\frac{1}{24}g_{ab}\left(3T_{cd}T^{cd}-T^2\right),
\end{equation}
where $T_{ab}$ is the energy-momentum tensor. The constants $\kappa^2$ and $\Lambda$ correspond to the gravitational coupling and cosmological constant on the brane. Throughout this work, we adopt relativistic units where $c = G = 1$, setting $\kappa^2 = 8\pi$.

While negative-tension branes are inherently unstable \cite{mypaper,bigrip}, frameworks incorporating variable brane tension have been extensively explored \cite{brane7}. The brane tension $\lambda$, bulk cosmological constant $\tilde{\Lambda}$, and bulk gravitational coupling $\tilde{\kappa}^2$ are linked to the brane parameters via the standard constraints:
\begin{eqnarray}
6\kappa^2 &=&\tilde{\kappa}^2\lambda,\\
2\Lambda &=& \kappa^2\lambda+\tilde{\kappa}^2 \tilde{\Lambda}.
\end{eqnarray}
A suggested alternative brane-world cosmology without dark radiation was proposed in \cite{brane14}. In this scenario, bulk black strings—higher-dimensional extensions of black holes—intersect the brane, leading to a brane universe with a “Swiss-cheese” (SC) structure \cite{brane15,nas333}. The presence of Schwarzschild black holes introduces important cosmological consequences and provides a potentially more realistic description than the conventional FLRW brane model \cite{br50}.

\quad Remarkably, these SC brane-world universes undergo eternal decelerating expansion, independently of the value of the cosmological constant $\Lambda$. For $\Lambda \leq 0$, the universe is characterized by a positive, monotonically decreasing energy density $\rho$ and a negative, increasingly large pressure $p$, eventually approaching the Einstein–Straus model at late times. The bulk black strings penetrate the brane and manifest themselves as Schwarzschild black holes from the four-dimensional brane perspective. The SC brane-world construction is fundamentally different from the original Swiss-cheese model introduced by Einstein and Straus describing compact objects embedded in an expanding cosmological background \cite{br51,br52}. In the Einstein–Straus model, a spherical comoving region is removed from an FLRW spacetime and replaced by a Schwarzschild region containing a central mass, thereby producing a spherical void within the cosmological background. This construction also leads to modifications of the luminosity–redshift relation \cite{br53}. The resulting geometry combines the large-scale homogeneous and isotropic cosmological evolution of the brane with the gravitational effects associated with black strings penetrating the bulk–brane system. In the flat FLRW regions, the cosmological equations are 
\begin{eqnarray}
\frac{\dot{a}^2}{a^2} &=& \frac{\Lambda}{3}+\frac{\kappa^2 \rho}{3}\left(1+\frac{\rho}{2\lambda}\right),  \label{cosm1}\\
\frac{\ddot{a}}{a} &=& \frac{\Lambda}{3}-\frac{\kappa^2 }{6}\left[\rho\left(1+\frac{2\rho}{\lambda}\right)+3p \left(1+\frac{\rho}{\lambda}\right) \right] .\label{cosm2}
\end{eqnarray}
GR is recovered for $\rho/\lambda\rightarrow 0$. The evolution of $\rho$ and $p$ in cosmological time $\tau$ is given as
\begin{eqnarray}
\frac{\rho}{\lambda} &=& -1 + \sqrt{1-\frac{2\Lambda}{\kappa^2 \lambda}+\frac{8}{3\kappa^2 \lambda \tau^2}},\\
\frac{p}{\lambda} &=& 1-\frac{4+3(\kappa^2 \lambda-2\Lambda )\tau^2}{(3\lambda)^{1/2} \kappa \tau \sqrt{8+3(\kappa^2 \lambda-2\Lambda )\tau^2}}.
\end{eqnarray}
A spatially flat dark energy-dominated SC brane-world universe, compatible with observational constraints, was obtained in \cite{mypaper}. Within the same framework, the possibility of a Big Rip singularity was subsequently investigated in \cite{bigrip}.

\quad The evolution of the energy density and pressure governed by (\ref{cosm1}) and (\ref{cosm2}) can be written in terms of the scale factor $a$. As previously demonstrated in Ref.~\cite{brane14}, the presence of the quadratic energy-density term leads to two possible branches of the energy density, corresponding to opposite signs, $\pm \rho$. Solving (\ref{cosm1}) and (\ref{cosm2}) separately for $\Lambda > 0$ and $\Lambda < 0$ therefore yields four distinct solutions, with two solutions associated with each sign of the cosmological constant.
\begin{equation} \label{rho1}
\rho_{1,2}=\frac{-2\pi a \lambda \pm \sqrt{\pi \lambda (4\pi \lambda a^2-\Lambda a^2+3 \dot{a}^2)}}{2\pi a},
\end{equation}
To determine the physically valid solution, we look at the prevailing sign of $\rho$ over the course of the entire cosmological cycle. In this model, the physical case corresponds to a positive $\rho$ alongside a negative cosmological constant \cite{role5,role10,dd4,bb,ma,NasrBamba}.

 Hence
\begin{eqnarray}
p&=&  \frac{\lambda(4\pi \rho\, a^2 + a^2 -2 \dot{a}^2- a\ddot{a}
)}{4\pi a^2 (\lambda + \rho)}~~~~~~~~~~~~~~ < 0, \label{p}\\
\rho&=&\frac{-2\pi a \lambda +\sqrt{\pi \lambda (4\pi \lambda a^2-\Lambda a^2+3 \dot{a}^2)}}{2\pi a} ~~> 0. ~~~~~ \lambda > 0 \label{rho}
\end{eqnarray}
where dots denote time derivatives. This expression reveals that negative tension branes are forbidden because they yield complex values. Incorporating a variable brane tension may offer a more realistic cosmic description \cite{brane7}, as its sign can vary according to the sign of the bulk cosmological constant.

\quad The present work aims to investigate the non-equilibrium thermodynamics of an expanding Swiss-cheese braneworld cosmology, with particular emphasis on quantum-corrected horizon entropy, entropy production, and the validity of the generalized second law of thermodynamics. In section 2, we introduce the Swiss-cheese braneworld framework and discuss the possible solutions. In section 3, We show that the standard equilibrium Clausius relation leads to the Bekenstein--Hawking area law, while the presence of corrected entropy terms naturally requires a non-equilibrium description with an additional entropy production contribution. In section 4, we derive the basic thermodynamical relations that will be revisited and modified for the non-equilibrium thermodynamic case. In section 5, we derive the corresponding entropy production rates and investigate the GSLT. In section 6, we study the matter bounce scenario and investigate the evolution of the matter, horizon, internal, and total entropy contributions together with the second derivative of the total entropy to assess the approach toward thermodynamic equilibrium. Finally, in section 7 we summarize main results and discusses possible extensions.

\section{A General Horizon Entropy function $S(A)$}

Instead of assuming a particular entropy formula, we consider a general entropy function of the apparent horizon area $S_h=S(A)$, $A=4\pi r_A^2$. For the effective fluid, the amount of energy crossing the apparent horizon during the time interval $dt$ is
\begin{equation}
\delta Q=A(\rho_{\rm eff}+p_{\rm eff})Hr_A\,dt.
\end{equation}
For a spatially flat universe $r_A=\frac{1}{H}$, this gives $Hr_A=1$. Hence,
\begin{equation} \label{dq}
\delta Q
=
A(\rho_{\rm eff}+p_{\rm eff})~dt.
\end{equation}

$T=\frac{1}{2\pi r_A}$ is the Hawking temperature associated with the apparent horizon. Since $S=S(A)$, then $dS=\frac{dS}{dA}dA$ and the Clausius relation $\delta Q=T\,dS$ gives $\delta Q=\frac{1}{2\pi r_A}\frac{dS}{dA}dA$. Using $dA=8\pi r_A\,dr_A$, then $TdS=4\frac{dS}{dA}dr_A$ and Eq. (\ref{dq}) becomes
\begin{equation}
A(\rho_{\rm eff}+p_{\rm eff})
=
4
\frac{dS}{dA}
\dot r_A.
\end{equation}

Therefore, $\frac{dS}{dA}=\frac{A(\rho_{\rm eff}+p_{\rm eff})}{4\dot r_A}$. Since $r_A=\frac{1}{H}$, then $\dot r_A=-\frac{\dot H}{H^2}$. Hence, 
\begin{equation}
\frac{dS}{dA}
=
-
\frac{
A(\rho_{\rm eff}+p_{\rm eff})H^2
}{
4\dot H
}.
\end{equation}

Using $A=\frac{4\pi}{H^2}$, we obtain the general formula

\begin{equation}
\label{eq:dSdA}
\frac{dS}{dA}
=
-\pi
\frac{
\rho_{\rm eff}+p_{\rm eff}
}{
\dot H
}.
\end{equation}

Using the identity $\frac{\ddot{a}}{a}=\dot{H}+H^2$ along with the SC Brane acceleration equation (\ref{cosm2}) gives
\begin{equation}\label{doth}
\dot{H}=- \frac{\kappa^2 }{2} (\rho +p)\left(1+\frac{\rho}{\lambda}\right),
\end{equation}
Comparing SC brane equations with the standard FLRW equations, we get
\begin{equation}
\rho_\mathrm{eff}=\rho \left(1+\frac{\rho}{2\lambda}\right), ~~p_{\text{eff}}=p+\frac{\rho^2}{2\lambda}+\frac{\rho p}{\lambda}.
\end{equation}

Hence, the effective fluid satisfies

\begin{equation} \label{eff}
\rho_{\rm eff}+p_{\rm eff}
=
(\rho+p)
\left(
1+\frac{\rho}{\lambda}
\right).
\end{equation}

Substituting in Eq.~(\ref{eq:dSdA}) gives

\begin{equation}
\frac{dS}{dA}
=
\frac{2\pi}{\kappa^2}.
\end{equation}

Integrating with respect to the horizon area,

\begin{equation}
S
=
\frac{2\pi}{\kappa^2}A+C.
\end{equation}

Using $\kappa^2=8\pi G$, the entropy becomes

\begin{equation}
S
=
\frac{A}{4G}
+C.
\end{equation}

Hence, If the effective fluid in SC braneworld universe is treated as the heat source along with assuming the standard Hawking temperature, the Clausius relation reconstructs exactly the Bekenstein-Hawking entropy with no correction terms. Consequently, the corrected entropy relation cannot be obtained from the standard Clausius relation together with the standard horizon temperature and the effective fluid description alone. To obtain correction terms, at least one of the assumptions must change: 1- The Clausius relation becomes a non-equilibrium relation, $\delta Q=TdS+TdS_i$ wher $dS_i$ is an entropy-production term. 2- The horizon temperature differs from the simple Hawking value $T\neq \frac{1}{2\pi r_A}$. 3- The heat flux contains additional contributions from bulk gravity or quantum effects that are not captured by $\rho_{ \rm eff}+p_{\rm eff}$. For Swiss-cheese braneworld, there are three possible sources of irreversibility: 1- Quadratic brane corrections $(\rho^2/\lambda)$ 2- Quantum corrections (the logarithmic entropy term) 3- Bulk-brane interactions (if energy exchange is allowed). Since our present model assumes Energy-Momentum conservation, the third source is absent. Therefore, the internal entropy production $dS_i$ should naturally be associated with the first two effects.

\section{Equilibrium Thermodynamics}

We start by deriving basic relations that will be revisited and modified for the non-equilibrium thermodynamic case. The homogeneous and isotropic universe is described by
\begin{equation}
ds^2
=
-dt^2
+
a^2(t)
\left(
\frac{dr^2}{1-kr^2}
+
r^2d\Omega^2
\right),
\end{equation}
Hayward rewrites the metric as 
\begin{equation}
ds^2
=
h_{ab}dx^adx^b
+
\tilde r^2d\Omega^2,
\end{equation}
where $x^a=(t,r)$. The metric $h_{ab}$ and its inverse are
\begin{equation}
h_{ab}=\mathrm{diag}\left(-1,\frac{a^2}{1-kr^2}\right) ~~h^{ab}=\mathrm{diag}\left(-1,\frac{1-kr^2}{a^2}\right).
\end{equation}
It is highly useful to rewrite the spatial radius as an areal (physical) radius $\tilde r(t,r)=a(t) r$ (Hayward's formalism). We then have $\partial_t\tilde r=\dot ar=Hr\,a=H\tilde r$ and $\partial_r\tilde r=a$. Hence
\begin{equation}
\partial \tilde r
=
(H\tilde r,a).
\end{equation}

We next determine the work density, energy-supply vector, and energy flux traversing the apparent horizon. These quantities appear in Hayward's unified first law and are independent of whether the thermodynamic description is equilibrium or non-equilibrium. We assume that there is no energy exchange between the brane and the bulk, so that the effective fluid is conserved:
\begin{equation}
\nabla^\mu
T_{\mu\nu}^{\rm eff}
=
0.
\end{equation}
Hence,
\begin{equation}
\dot\rho_{\rm eff}
+
3H
(\rho_{\rm eff}+p_{\rm eff})
=
0.
\end{equation}
Substituting the effective quantities,

\begin{equation}
\dot\rho \left(1+\frac{\rho}{\lambda}\right)+3H (\rho+p)\left(1+\frac{\rho}{\lambda}\right)=0
\end{equation}
Since $\left(1+\frac{\rho}{\lambda}\right) > 0$, we recover
\begin{equation}
\dot\rho
+
3H(\rho+p)
=
0.
\end{equation}
This confirms that ordinary matter is conserved on the brane. 
The work density $W=-\frac12 T^{ab}h_{ab}=\frac12 (\rho_{\rm eff}-p_{\rm eff})$ is given by
\begin{equation}
W=\frac12 \left(\rho-p -\frac{\rho p}{\lambda} \right).
\end{equation}
For dust $p=0$ gives $W=\frac{\rho}{2}$, for radiation $p=\frac{\rho}{3}$ gives $W=\frac{\rho}{3}-\frac{\rho^2}{6\lambda}$, and for vacuum energy $p=-\rho$ gives $W=\rho+\frac{\rho^2}{2\lambda}$. The Energy-supply vector is
\begin{equation}
\Psi_a
=
T_a^{\ b}
\partial_b r
+
W
\partial_a r.
\end{equation}

Since $T_t^{~t}=-\rho_{\rm eff}$, $\partial_t r_A=H r_A$, and $\partial_r r_A=a$, we obtain $\Psi_t=-\frac{1}{2} (\rho_{\rm eff}+p_{\rm eff})H r_A$. Similarly, $T_r^{~r}=p_{\rm eff}$. Hence $\Psi_r=a p_{\rm eff}+a W$. Using $W$ definition gives $\Psi_r=\frac12 (\rho_{\rm eff}+p_{\rm eff})a$. Finally,
\begin{equation}
\Psi_a
=
\left(
-\frac12
(\rho_{\rm eff}+p_{\rm eff})
Hr_A,
\;
\frac12
(\rho_{\rm eff}+p_{\rm eff})
a
\right).
\end{equation}
Substituting $\rho_{\rm eff}$ and $p_{\rm eff}$ for the Swiss-cheese brane,
\begin{equation}
\Psi_a
=
\frac12
(\rho+p)
\left(
1+\frac{\rho}{\lambda}
\right)
(-Hr_A,a).
\end{equation}

Hayward's unified first law is
\begin{equation}
dE
=
A\Psi
+
WdV.
\end{equation}
 $dE$ is the change of Misner-Sharp energy, $A \Psi$ is the energy flux, $W dV$ is the work done by the effective fluid. The Clausius relation $\delta Q = T_h dS_h$ is assumed to hold exactly for a universe evolves in thermodynamic equilibrium. The Misner-Sharp energy enclosed within the apparent horizon is $E=\rho_{\rm eff}V$ where $V=\frac{4\pi}{3}r_A^3$. Therefore,
\begin{equation}
E
=
\frac{4\pi}{3}r_A^3\rho_{\rm eff}.
\end{equation}
Taking the total differential,
\begin{equation}
dE
=
4\pi r_A^2\rho_{\rm eff}\,dr_A
+
\frac{4\pi}{3}r_A^3\,d\rho_{\rm eff}.
\end{equation}
Using $d\rho_{\rm eff}=\dot\rho_{\rm eff}dt$ gives
\begin{equation}
dE
=
4\pi r_A^2\rho_{\rm eff}\dot r_A dt
+
\frac{4\pi}{3}r_A^3
\dot\rho_{\rm eff}dt.
\end{equation}
Since
\begin{equation}
\dot\rho_{\rm eff}
=
-3H
(\rho_{\rm eff}+p_{\rm eff}),
\end{equation}
we obtain
\begin{equation}
dE
=
4\pi r_A^2\rho_{\rm eff}\dot r_A dt
-
4\pi Hr_A^3
(\rho_{\rm eff}+p_{\rm eff})dt.
\end{equation}
Since the work density definition along with the differential of the volume $dV=4\pi r_A^2dr_A$. we have,
\begin{equation}
WdV
=
2\pi r_A^2
(\rho_{\rm eff}-p_{\rm eff})
dr_A.
\end{equation}
Substituting in $A\Psi=dE-WdV$,
\begin{align}
A\Psi
=&
4\pi r_A^2\rho_{\rm eff}dr_A
-
4\pi Hr_A^3
(\rho_{\rm eff}+p_{\rm eff})dt
\nonumber\\
&
-
2\pi r_A^2
(\rho_{\rm eff eff}-p_{\rm eff})dr_A.
\end{align}
collecting the $dr_A$ terms,
\begin{equation} 
A\Psi=2\pi r_A^2(\rho_{\rm eff}+p_{\rm eff})dr_A-4\pi Hr_A^3(\rho_{\rm eff}+p_{\rm eff})dt
\end{equation}
Factorizing,
\begin{equation} \label{Apsi}
A\Psi
=
2\pi
(\rho_{\rm eff}+p_{\rm eff})
\left(
r_A^2dr_A
-
2Hr_A^3dt
\right).
\end{equation}
The heat crossing the horizon is defined by $\delta Q=-A\Psi$. Using (\ref{Apsi}),
\begin{equation}
\delta Q
=
-2\pi
(\rho_{\rm eff}+p_{\rm eff})
\left(
r_A^2dr_A
-
2Hr_A^3dt
\right).
\end{equation}
Along the apparent horizon we have $dr_A=\dot r_Adt$. Thus,
\begin{equation}
\delta Q
=
-2\pi
(\rho_{\rm eff}+p_{\rm eff})
\left(
r_A^2\dot r_A
-
2Hr_A^3
\right)dt.
\end{equation}

However, projecting the unified first law along the horizon generator eliminates the work contribution, leaving $\delta Q=A(\rho_{\rm eff}+p_{\rm eff})Hr_Adt$ \cite{cai2}. Since $Hr_A=1$ (k=0), we finally obtain the standard heat flow across the apparent horizon in a flat FLRW universe
\begin{equation}
\delta Q
=
4\pi r_A^2
(\rho_{\rm eff}+p_{\rm eff})dt.
\end{equation}
Making use of Clausius relation $\delta Q=T_hdS_h$,  
\begin{equation}
4\pi r_A^2
(\rho_{\rm eff}+p_{\rm eff})dt
=
\frac1{2\pi r_A}dS_h.
\end{equation}
Therefore,
\begin{equation}
dS_h
=
8\pi^2
r_A^3
(\rho_{\rm eff}+p_{\rm eff})dt.
\end{equation}

\section{Non-equilibrium Thermodynamics of the Swiss-Cheese Braneworld}

In irreversible thermodynamics, the total entropy is given as 

\begin{equation}
S_{tot}=S_h+S_i
\end{equation}
where $S_h$ is the reversible horizon entropy, and $S_i$ is the internal irreversible entropy production. The generalized Clausius relation becomes
\begin{equation} \label{clau}
\delta Q=T_h dS_{tot}=T_h (dS_h+ dS_i),
\end{equation}
The heat flux is unchanged because it is determined solely by the effective energy-momentum tensor,
\begin{equation}
\delta Q
=
A
(\rho_{\rm eff}+p_{\rm eff})
Hr_A\,dt.
\end{equation}
For the flat universe $Hr_A=1$,
\begin{equation}
\delta Q
=
4\pi r_A^2
(\rho_{\rm eff}+p_{\rm eff})dt.
\end{equation}
Which is identical to the equilibrium case, the heat crossing the horizon does not change. The Hawking temperature $T_h$ also remains unchanged,
thus
\begin{equation}
T_hdS_h=\frac1{2\pi r_A}dS_h~,~~~   T_h dS_i=\frac1{2\pi r_A}dS_i.
\end{equation}
Substituting the previous expressions into the generalized Clausius relation gives
\begin{equation}
4\pi r_A^2
(\rho_{\rm eff}+p_{\rm eff})dt
=
\frac1{2\pi r_A}
\left(
dS_h+dS_i
\right).
\end{equation}
Hence,
\begin{equation}
dS_h+dS_i
=
8\pi^2r_A^3
(\rho_{\rm eff}+p_{\rm eff})dt.
\end{equation}
This equation replaces the equilibrium entropy relation obtained in previous section. The reversible contribution is still assumed to satisfy the Bekenstein-Hawking relation,
\begin{equation} \label{hawkent}
S_h^{({\rm rev})}
=
\frac{A}{4G}.
\end{equation}
Hence $dS_h=\frac1{4G}dA$. Using $A=4\pi r_A^2$ gives $dA=8\pi r_Adr_A$. Therefore,
\begin{equation}
dS_h
=
\frac{2\pi}{G}
r_Adr_A.
\end{equation}
Subtracting the reversible contribution from the entropy balance yields
\begin{equation}
dS_i
=
8\pi^2r_A^3
(\rho_{\rm eff}+p_{\rm eff})dt
-
\frac{2\pi}{G}
r_Adr_A.
\end{equation}
This is the general entropy-production equation for the Swiss-cheese braneworld. No explicit form of $dS_i$ has been assumed. Using $r_A=\frac1H$ and $dr_A=-\frac{\dot H}{H^2}dt$, we obtain
\begin{eqnarray}
\frac{dS_i}{dt}
&=&
\frac{8\pi^2}{H^3}
(\rho_{\rm eff}+p_{\rm eff})
+
\frac{2\pi}{GH^3}
\dot H, \nonumber \\
&=& \frac{8\pi^2}{H^3}
(\rho+p)
\left(
1+\frac{\rho}{\lambda}
\right)
+
\frac{2\pi}{GH^3}
\dot H
\end{eqnarray}

Substituting $\dot{H}$ from Eq. (\ref{doth}) will cancel the two terms leading to zero entropy production $d S_i=0$, which can be considered as a direct result of keeping the same equilibrium assumptions unchanged. The next step is identifying the physical mechanism (quantum corrections, modified heat flux, or a non-equilibrium horizon description) that breaks the cancellation and produces the entropy correction. We will keep the horizon entropy completely general. Instead of assuming $S_h=\frac{A}{4G}$, we let $S_h=S(A)$ as assumed before. Then $dS_h=S'(A)\,dA$ where $S'(A) \equiv \frac{dS}{dA}$. Substituting $T_h=\frac{1}{2\pi r_A}$ and $dS_h=S'(A)\,dA$ in the non-equilibrium Clausius relation (\ref{clau}) gives
\begin{eqnarray}
\delta Q &=& \frac1{2\pi r_A}\left[S'(A)dA+dS_i\right],~~~~~~~~~~dA=8\pi r_Adr_A \nonumber \\  
&=& 4S'(A)dr_A+\frac{dS_i}{2\pi r_A}.
\end{eqnarray}

The energy crossing the horizon remains
\begin{equation}
\delta Q
=
4\pi r_A^2
(\rho_{\rm eff}+p_{\rm eff})dt.
\end{equation}
Equating both expressions,
\begin{equation}
4\pi r_A^2
(\rho_{\rm eff}+p_{\rm eff})dt
=
4S'(A)dr_A
+
\frac{dS_i}{2\pi r_A}.
\end{equation}
Hence,
\begin{equation}
4\pi r_A^2
(\rho_{\rm eff}+p_{\rm eff})
=
4S'(A)\dot r_A
+
\frac{\dot S_i}{2\pi r_A}.
\end{equation}
Using  $r_A=1/H$, $\dot r_A=-\frac{\dot H}{H^2}$, and Multiplying by $H^2$ we get

\begin{equation}
4\pi
(\rho_{\rm eff}+p_{\rm eff})
=
-
4S'(A)\dot H
+
\frac{H^3}{2\pi}\dot S_i.
\end{equation}
Hence
\begin{equation}
\dot S_i
=
\frac{2\pi}{H^3}
\left[
4\pi
(\rho_{\rm eff}+p_{\rm eff})
+
4
S'(A)
\dot H
\right].
\end{equation}
Finally, using (\ref{doth}) and (\ref{eff}) for the Swiss-cheese braneworld 
\begin{equation}
\dot S_i
=
\frac{8\pi}{H^3}
(\rho+p)
\left(1+\frac{\rho}{\lambda}\right)
\left[
\pi
-
\frac{\kappa^2}{2}
S'(A)
\right].
\end{equation}
The above entropy production does not vanish identically because the reversible entropy is no longer restricted to the Bekenstein-Hawking form. Instead, it is determined by the corrected entropy functional. The modified entropy, such as the quantum-corrected entropy relation ( \ref{ent}), an possibly be interpreted as the sum of the equilibrium Bekenstein-Hawking entropy (\ref{hawkent}) plus an irreversible entropy required by the modified gravity. Substituting 
\begin{equation} \label{Sdash}
S_h'(A_h)
=
\frac{1}{4G}
+\frac{\alpha}{A_h}
-\frac{4\beta G}{A_h^2},
\end{equation}
we obtain
\begin{equation}
\dot S_i
=
\frac{8\pi}{H^3}
(\rho+p)
\left(1+\frac{\rho}{\lambda}\right)
\left[
\pi
-\frac{\kappa^2}{2}
\left(
\frac{1}{4G}
+\frac{\alpha}{A_h}
-\frac{4\beta G}{A_h^2}
\right)
\right].
\end{equation}
Using $\kappa^2=8\pi G$, the classical terms cancel $\pi-\frac{\kappa^2}{8G}=0$. Thus

\begin{equation}
\dot S_i
=
\frac{8\pi}{H^3}
(\rho+p)
\left(1+\frac{\rho}{\lambda}\right)
\left[
-\frac{\kappa^2\alpha}{2A_h}
+\frac{2\kappa^2\beta G}{A_h^2}
\right].
\end{equation}
Using $A=4\pi/H^2$, this becomes

\begin{equation}
\dot S_i
=
\kappa^2
(\rho+p)
\left(1+\frac{\rho}{\lambda}\right)
\left(
-\frac{\alpha}{H}
+\frac{\beta GH}{\pi}
\right).
\end{equation}

Using (\ref{doth}), the internal entropy production can be expressed in terms of $H$

\begin{equation} \label{inent}
\dot S_i
=
2\dot H
\left(
\frac{\alpha}{H}
-\frac{\beta GH}{\pi}
\right).
\end{equation}

Since, $\dot S_h=\frac{dS_h}{dA_h}\frac{dA_h}{dt}$. Using (\ref{Sdash}) we have
\begin{equation}
\dot S_h
=
\left(
\frac{1}{4G}
+
\frac{\alpha}{A_h}
-
\frac{4\beta G}{A_h^2}
\right)
\dot A_h.
\end{equation}
Differentiating the horizon area $A_h=4\pi r_A^2$ with respect to time gives $\dot A_h=8\pi r_A\dot r_A$. Hence,

\begin{equation}
\dot S_h
=
8\pi r_A\dot r_A
\left(
\frac{1}{4G}
+
\frac{\alpha}{A_h}
-
\frac{4\beta G}{A_h^2}
\right).
\end{equation}

For a spatially flat FLRW universe $r_A=\frac{1}{H}$, then $\dot r_A=-\frac{\dot H}{H^2}$ and $8\pi r_A\dot r_A=-\frac{8\pi\dot H}{H^3}$. The area also can be expressed in terms of $H$ as $A_h=4\pi r_A^2=\frac{4\pi}{H^2}$. We get

\begin{equation}
\dot S_h
=
-\frac{8\pi\dot H}{H^3}
\left[
\frac{1}{4G}
+
\frac{\alpha H^2}{4\pi}
-
\frac{\beta G H^4}{4\pi^2}
\right].
\end{equation}

The total entropy now is $S_{\rm tot}=S_m+S_h+S_i$. Therefore,

\begin{eqnarray}
\dot S_{\rm tot}&=&\dot S_m+\dot S_h+\dot S_i \\   \nonumber
&=& \frac{4\pi}{TH^2}
(\rho+p)
\left[
\frac{\kappa^2}{2H^2}
(\rho+p)
\left(1+\frac{\rho}{\lambda}\right)-1
\right]
-
\frac{8\pi\dot H}{H^3}
\left[
\frac{1}{4G}
+\frac{\alpha H^2}{4\pi}
-\frac{\beta GH^4}{4\pi^2}
\right]
\\    \nonumber
&+&
\kappa^2(\rho+p)
\left(1+\frac{\rho}{\lambda}\right)
\left(
-\frac{\alpha}{H}
+\frac{\beta GH}{\pi}
\right).
\end{eqnarray}
where the matter entropy $S_\mathrm{m}$ inside the horizon is obtained from the Gibbs relation $T_\mathrm{m} dS_\mathrm{m}=dE+p dV$ assuming thermal equilibrium between the apparent horizon and the matter content $T_\mathrm{m}=T$ With $E=\rho_{\text{eff}} V$ and $V=4\pi r_\mathrm{A}/3$. The generalized second law requires $\dot S_{\rm tot}\geq0$.

\section{Matter-bounce model}
\quad Among the different classes of bouncing cosmological models, the Matter Bounce Scenario (MBS) has attracted significant attention \cite{bounc11,bounc12}. In this framework, the universe undergoes an approximately matter-dominated contracting phase before smoothly transitioning through a nonsingular bounce. At the bounce, the entire universe is assumed to have been in causal contact, thereby providing a natural resolution to the horizon problem. The bounce is subsequently followed by a conventional expanding phase that reproduces the expected behavior of the standard cosmological picture. Various aspects and implications of the MBS have been investigated in Ref.~\cite{bounc5}. The MBS can be described by the following choice of scale factor for $n=\frac{1}{3}$ \cite{bounc5} 
\begin{equation} \label{scalef}
a(\tau)=\left(\eta \tau^2+1\right)^n
\end{equation}
With $\eta>0$, the bounce occurs at $\tau=0$, where $a(0)=1$. For the Hubble parameter we get,
\begin{equation} \label{hdoth}
H= \frac{2n ~\eta \tau}{ \eta \tau^2+1},\; \dot{H}=\frac{2n ~\eta(1-\eta \tau^2)}{(\eta \tau^2+1)^2},
\end{equation}
Hence, $H(0)=0$ and $\dot{H}(0)=2n \eta >0$ (bounce condition). The Ricci and Kretschmann scalars 
\begin{eqnarray} \label{RicciK}
R&=&6(\dot{H}+2H^2+\frac{k}{a^2}),\\  
K&=&12\left[(\dot{H}+H^2)^2+\left(H^2+\frac{k}{a^2}\right)^2\right].
\end{eqnarray}
Both $R$ and $K$ are finite at the bounce where $R(0)=12 ~n \eta $ and $K(0)=12~(2n \eta)^2$. The deceleration and jerk parameters are
\begin{equation} 
q=\frac{1}{2}-\frac{3}{2 \eta \tau^2},~~ j=1+\frac{3}{2 \eta \tau^2}.
\end{equation}
As $t \rightarrow 0$, $q \rightarrow -\infty$. Far from the bounce, as ($t \rightarrow \infty$), the deceleration parameter approaches $\frac{1}{2}$, corresponding to the standard matter-dominated universe. The jerk parameter diverges as $\tau \rightarrow 0$, while it approaches $1$ in the asymptotic limit $\tau \rightarrow \infty$ recovering the standard value for a matter-dominated FLRW universe. For the scale factor $a(\tau)=\left(1+\eta \tau^2\right)^{1/3}$, $H(0)=0$ and $\dot H(0)=\frac{2\eta}{3}>0$ at the bounce. The evolution of the effective equation of state parameter provides an additional indication of the asymptotic behavior of the model. Away from the bounce, \(w_{\rm eff}\) remains very close to \(-1\), indicating that the effective cosmological fluid approaches a dark energy-like, or cosmological constant-like, regime. The significant deviation from \(w_{\rm eff}=-1\) is confined primarily to the vicinity of the bounce, where the high-energy braneworld corrections become important and govern the nonsingular transition. Thus, rather than being dark-energy dominated throughout its entire evolution, the model naturally evolves from a high energy bounce regime into an asymptotically dark energy-like phase on the expanding branch \cite{caii,tempo,nas111}. Using (\ref{doth}) for the SC brane,
\begin{equation}
(\rho+p)
\left(1+\frac{\rho}{\lambda}\right)
=
-\frac{2\dot H}{\kappa^2}.
\end{equation}
For the matter bounce,
\begin{equation}
(\rho+p)
\left(1+\frac{\rho}{\lambda}\right)
=
-\frac{4\eta(1-\eta \tau^2)}
{3\kappa^2(1+\eta \tau^2)^2}.
\end{equation}
Horizon radius and area are 
\begin{equation}
r_A
=
\frac{3(1+\eta \tau^2)}
{2\eta \tau}, ~~~ A_h
=
\frac{9\pi(1+\eta \tau^2)^2}
{\eta^2 \tau^2}.
\end{equation}
The internal entropy production for the matter bounce can be obtained using (\ref{inent}) which gives
\begin{equation}
\dot S_i
=
\frac{2\alpha(1-\eta \tau^2)}
{\tau(1+\eta \tau^2)}
-
\frac{8\beta G \eta^2 \tau(1-\eta \tau^2)}
{9\pi(1+\eta \tau^2)^3}.
\end{equation}

Figure (1) shows the evolution of $\dot S_m$, $\dot S_h$, $\dot S_i$, and $\dot S_{\rm tot}$ for the MB model. Throughout the expanding phase, the MB-SC braneworld model respects the GSLT by maintaining a positive total entropy production rate ($\dot S_{tot} > 0$). As long as $\dot{S}_{\text{tot}} \ge 0$ is preserved across all evolutionary stages, the horizon contribution effectively counteracts the geometric dip to keep the GSLT intact.

\quad At the cosmic bounce point, the second time derivative of total entropy, $\ddot{S}_{\text{tot}}$, reaches a positive maximum, forming a symmetric downward-opening parabola. This peak reflects an accelerated rate of entropy creation driven by modified gravity effects and strong non-equilibrium conditions. Moving away from $t = 0$ in either direction, contracting ($t < 0$) or expanding ($t > 0$), $\ddot{S}_{\text{tot}}$ transitions smoothly into negative values. During late-time expansion ($t \gg 0$), $\ddot{S}_{\text{tot}} < 0$ indicates that entropy generation is slowing down, confirming that the universe is gradually relaxing toward thermodynamic equilibrium in accordance with the maximum entropy principle.

\begin{figure}[H]
  \centering         
	  \subfigure[$\rho > 0$ and $p < 0$]{\label{re3}\includegraphics[width=0.3\textwidth]{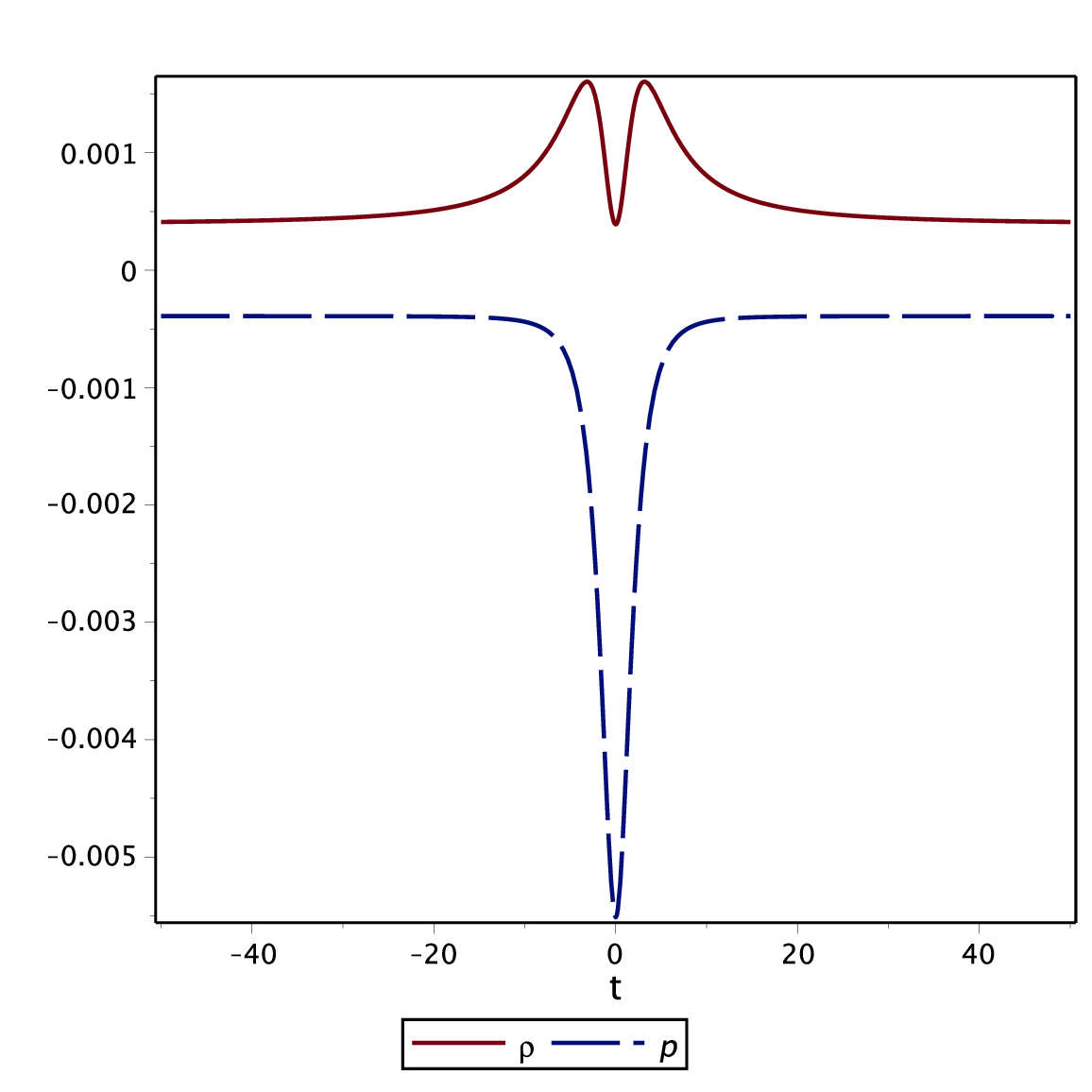}} 
		 \hspace{0.3cm}
			\subfigure[$w$]{\label{0yy0}\includegraphics[width=0.3\textwidth]{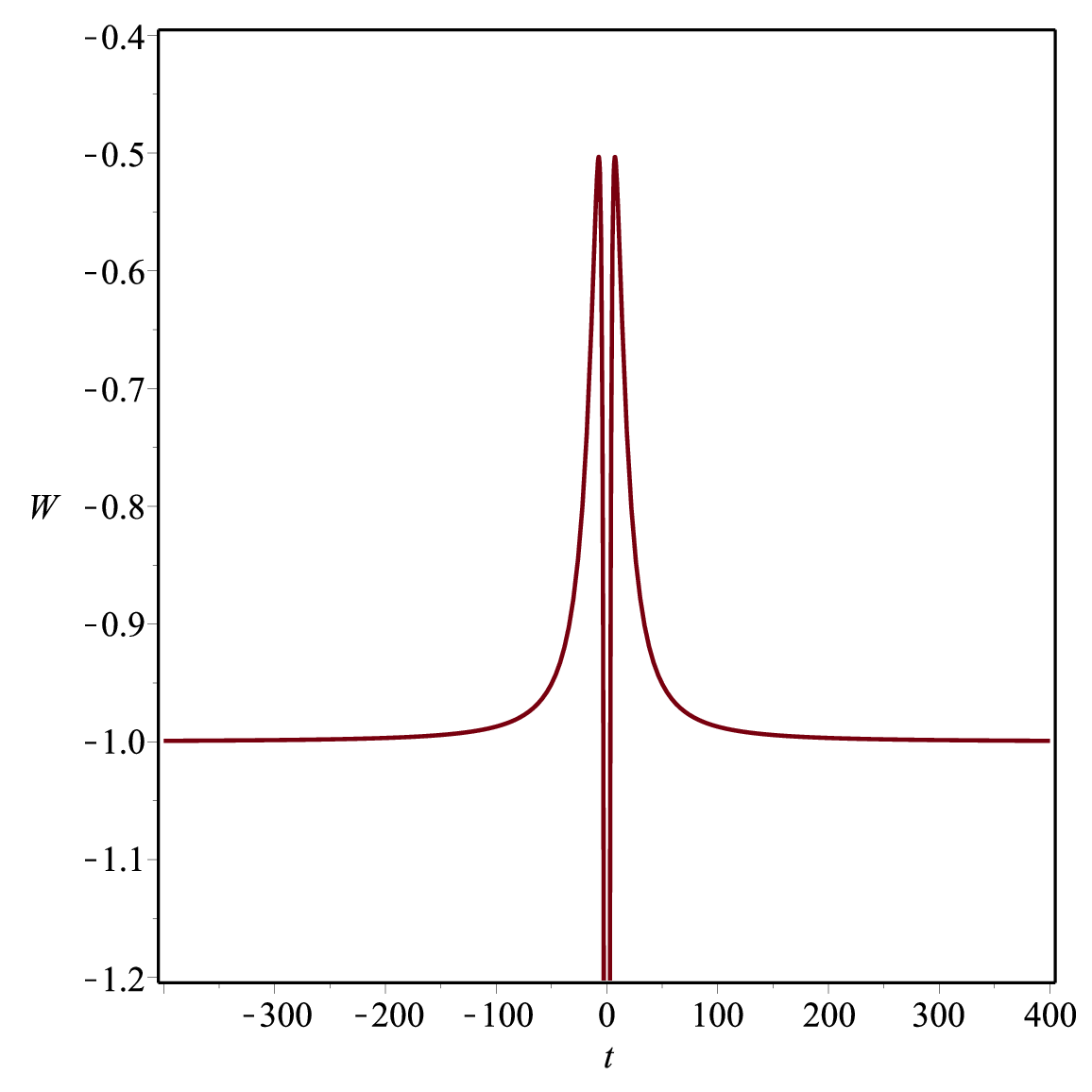}}
		\hspace{0.3cm}
			  \subfigure[$K$]{\label{lo3}\includegraphics[width=0.3\textwidth]{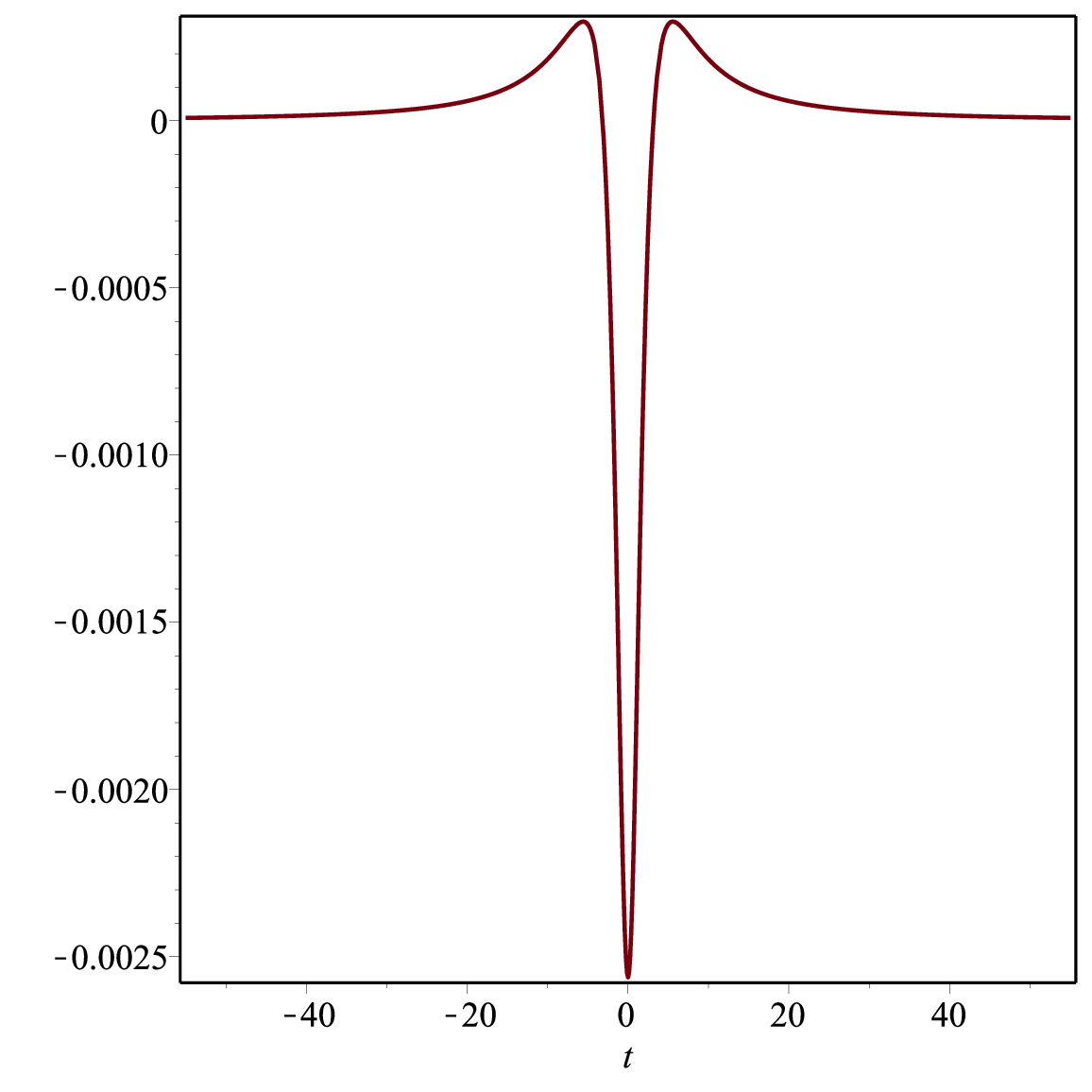}} \\
		 \subfigure[$\dot S_m$ and $\dot S_h$]{\label{Fw}\includegraphics[width=0.3\textwidth]{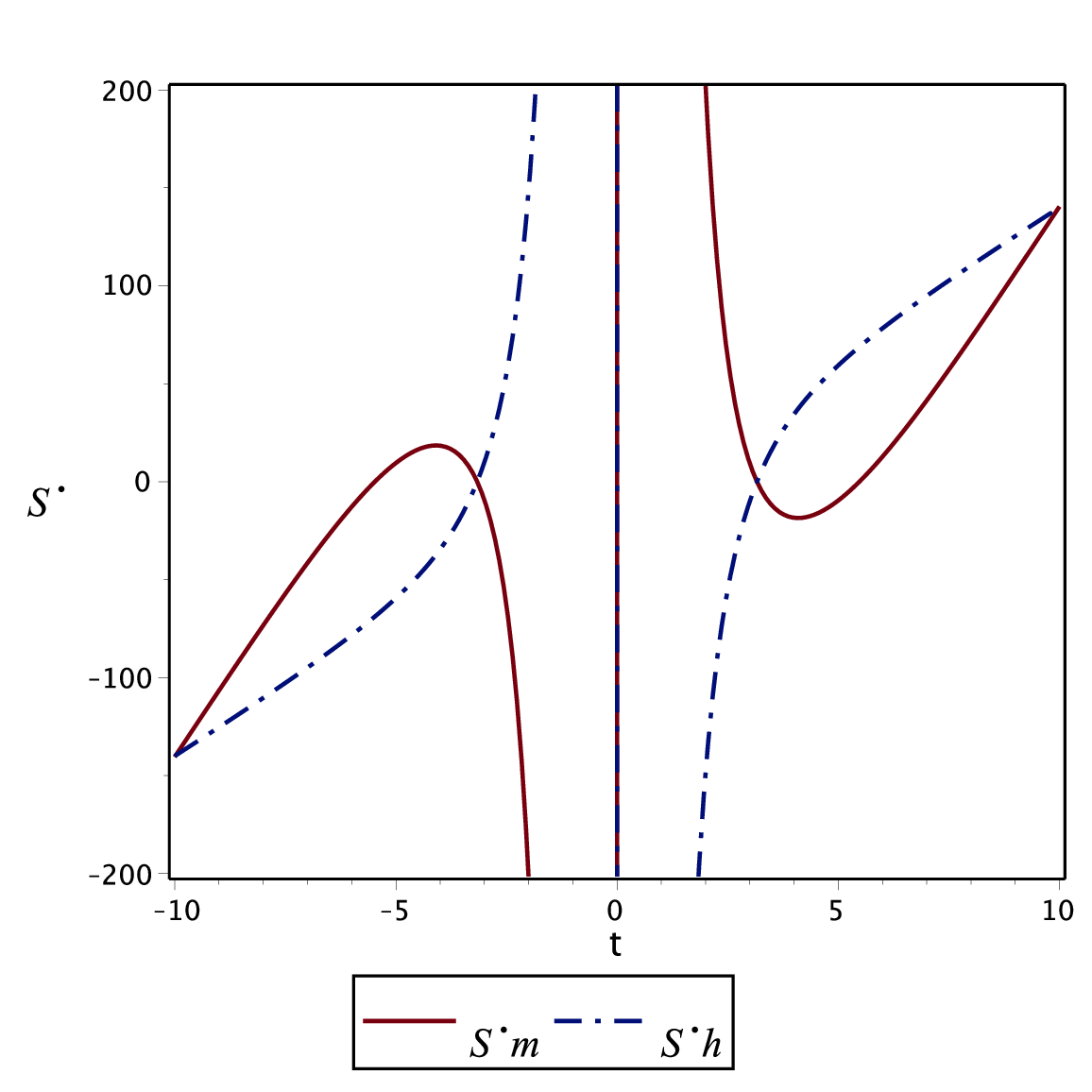}}
		 \hspace{0.3cm}
		 \subfigure[$\dot S_i$]{\label{add}\includegraphics[width=0.3\textwidth]{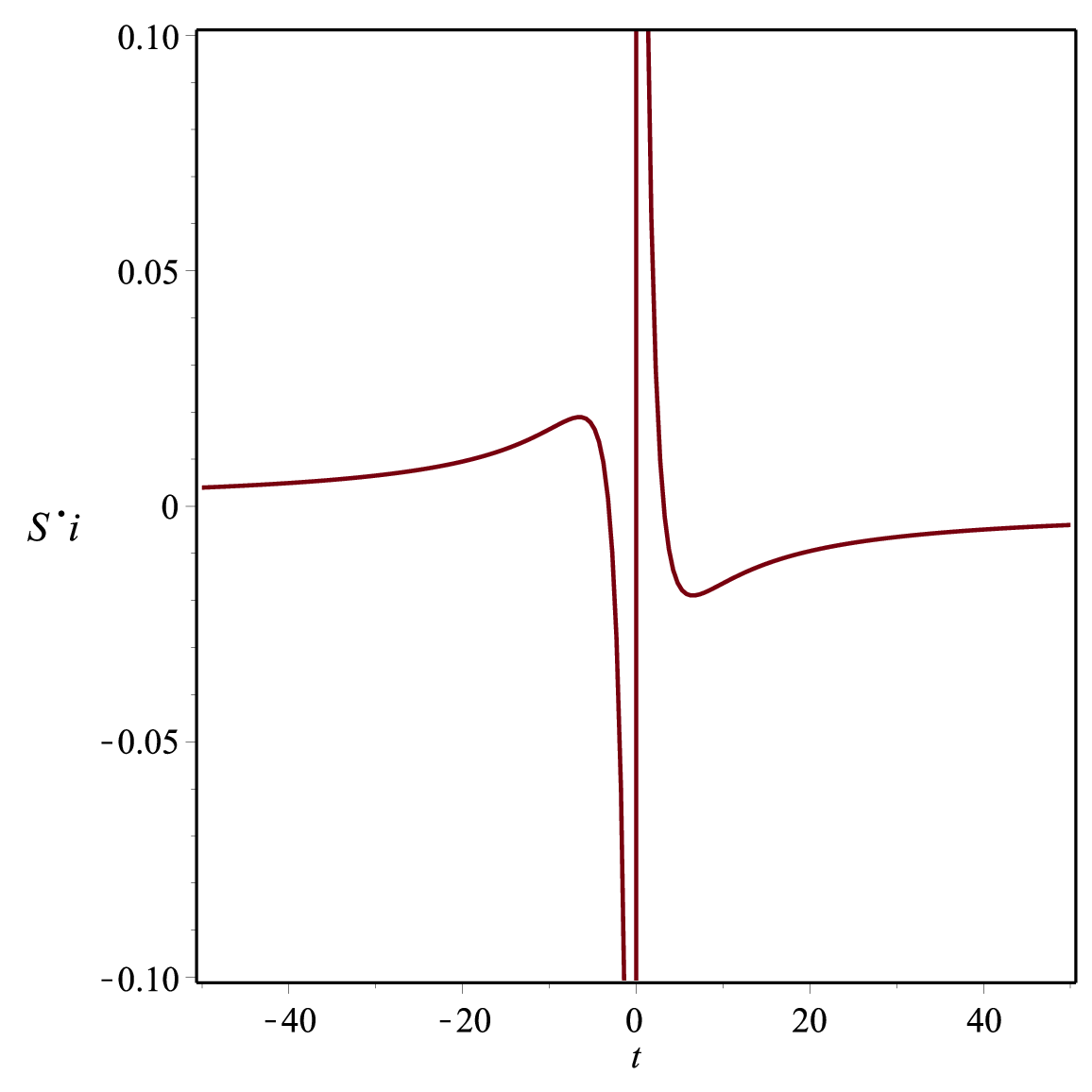}}
		 \hspace{0.3cm}
			\subfigure[$\dot S_{total}$]{\label{ass}\includegraphics[width=0.3\textwidth]{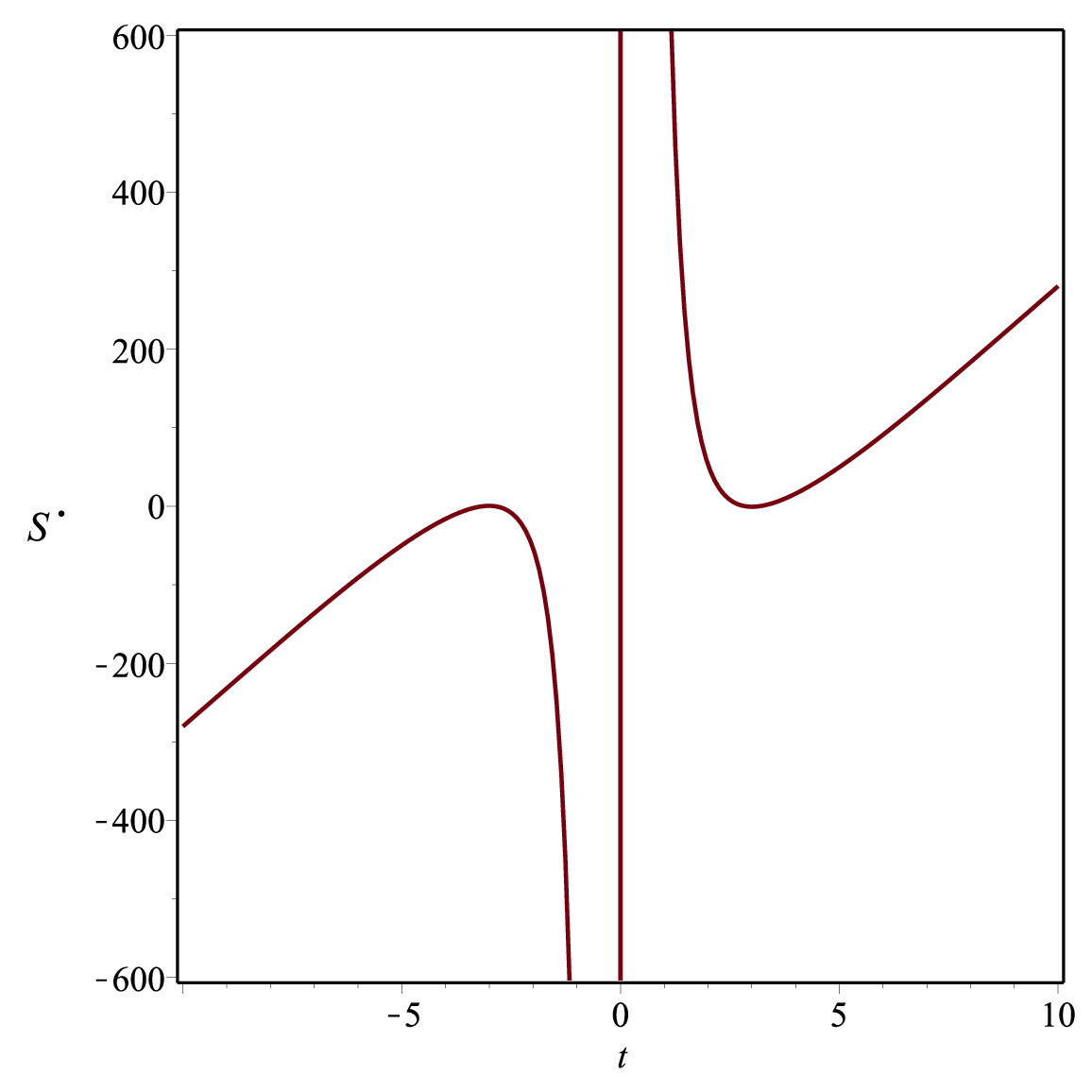}}
  \caption{ (a) Physical evolution of $\rho$ and $p$. In the Standard Model, positive pressure dominates during the early universe (as $z \to \infty$), whereas negative pressure drives the cosmic expansion during its late-time accelerated phase. (b) $w \approx -1$ with quintessence and quintom-like behavior appears only in a narrow region around the bounce. (c) The kinetic term is positive except in the neighborhood of the bounce which implies the absence of ghost-like behavior. (d) $\dot S_m$ and $\dot S_h$ are always positive in the expanding phase except near the bounce for $\dot S_h$. (e) Irreversible entropy production rate $\dot S_i$ verses cosmic time, it is positive in the contracting phase $t<0$ and negative in the expanding phase $t>0$ except around the bounce. Since we are not considering bulk viscosity or particle creation rate, this is likely a signature of high-energy geometrical corrections near the bounce. (f) The total entropy production rate $\dot S_{tot} >0$ in the expanding phase, satisfying the GSLT for the MB-SC braneworld model. As long as $\dot{S}_{\text{tot}} \ge 0$ is strictly maintained across the entire evolution, the model successfully satisfies the GSLT, with the horizon term compensating for the geometric dip.}
  \label{fig:cassimir55}
\end{figure}

\begin{figure}[H]
    \centering
    \includegraphics[width=0.45\textwidth]{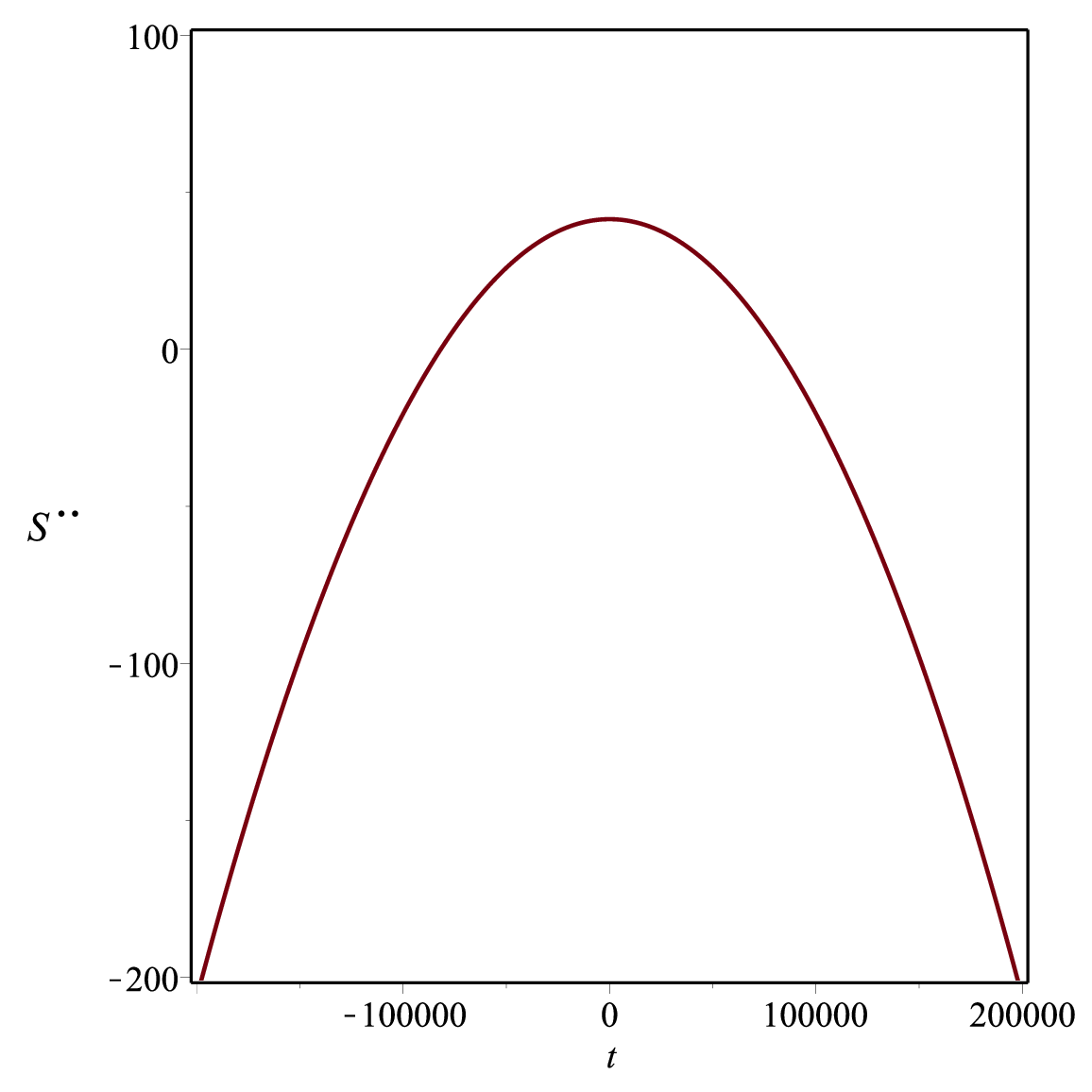}
    \caption{$\ddot{S_{tot}}$: The second time derivative of the total entropy $\ddot{S}_{\text{tot}}$ exhibits a symmetric downward-parabolic behavior centered at the bounce point. Near the bounce, $\ddot{S}_{\text{tot}}$ attains a positive peak, signaling an accelerated rate of entropy production driven by strong non-equilibrium effects and modified gravity dynamics during the cosmic bounce. As the universe evolves away from $t = 0$ in both the contracting ($t < 0$) and expanding ($t > 0$) phases, $\ddot{S}_{\text{tot}}$ smoothly shifts to negative values. In particular, during late-time expansion ($t \gg 0$), $\ddot{S}_{\text{tot}} < 0$ confirms that the entropy growth decelerates, demonstrating the universe's ultimate relaxation toward thermodynamic equilibrium in full agreement with the maximum entropy principle.}
    \label{fij}
\end{figure}


\section{Conclusion} \label{conclusion}

In this work, we have investigated the thermodynamic behavior of a Swiss-Cheese braneworld universe within a non-equilibrium framework, with particular emphasis on quantum-corrected horizon entropy and the generalized second law of thermodynamics. By combining the effective SC braneworld description with Hayward’s unified first law, we established a connection between the modified cosmological dynamics and the thermodynamics of the apparent horizon. We first considered a general entropy function \(S_h=S(A_h)\) and showed that, when the effective braneworld fluid is treated as the heat source and the standard Hawking temperature is adopted, the equilibrium Clausius relation uniquely reproduces the Bekenstein–Hawking entropy, $S_h=\frac{A_h}{4G}$. Thus, the quadratic \(\rho^2/\lambda\) correction in the SC braneworld Friedmann equations does not by itself generate a correction to the horizon entropy within the equilibrium description. This result indicates that a modified entropy-area relation requires an extension of the equilibrium Clausius relation, for example through an entropy production term, a modified horizon temperature, or additional contributions to the horizon heat flux. We subsequently formulated the horizon thermodynamics in a non-equilibrium framework, $\delta Q=T_h(dS_h+dS_i)$, where \(S_i\) represents the internal entropy associated with irreversible processes. For a general entropy functional, we derived the corresponding entropy production rate. 
We then applied the formalism to the matter-bounce solution which describes a nonsingular transition from contraction to expansion. The analysis shows that the different entropy contributions have distinct behaviors around the bounce. In particular, the internal entropy-production rate changes sign between the contracting and expanding phases, while the horizon and matter contributions compensate for the non-equilibrium behavior near the bounce. Most importantly, the total entropy production rate remains positive throughout the expanding phase, indicating that the GSLT is respected by the matter-bounce SC braneworld model.

The second derivative of the total entropy provides an additional thermodynamic diagnostic. Its late-time negative behavior indicates that the entropy growth slows down as the universe evolves away from the bounce. Consequently, the expanding universe approaches a state of thermodynamic equilibrium in accordance with the maximum entropy principle. The combination $\dot S_{\rm tot}\geq0, \qquad \ddot S_{\rm tot}<0$ at late times provides a consistent thermodynamic picture of the evolution. Throughout the expanding phase, the MB-SC braneworld model respects the Generalized Second Law of Thermodynamics by maintaining a positive total entropy production rate ($\dot S_{tot} > 0$). As long as $\dot{S}_{\text{tot}} \ge 0$ is preserved across all evolutionary stages, the horizon contribution effectively counteracts the geometric dip to keep the GSLT intact. 

The behavior of the effective equation of state parameter, $w_{\rm eff}$, further highlights the asymptotic dynamics of the model. Far from the bounce point, $w_{\rm eff}$ remains close to $-1$, signaling that the effective cosmic fluid approaches a dark energy- or cosmological constant-dominated regime. The departures from $w_{\rm eff} = -1$ occurs strictly near the bounce, where high energy braneworld corrections drive the non-singular cosmic transition. Consequently, instead of requiring dark energy dominance through all epochs, the model smoothly evolves from a high energy bounce into an asymptotic dark energy phase during late-time expansion.

Our results show that the non-equilibrium formulation is essential for consistently incorporating quantum-corrected horizon entropy into SC braneworld cosmology. A natural extension of the present work is to investigate linear cosmological perturbations and the corresponding thermodynamic constraints.




\end{document}